\documentclass[sigconf]{acmart}
\AtBeginDocument{
  \providecommand\BibTeX{{
    \normalfont B\kern-0.5em{\scshape i\kern-0.25em b}\kern-0.8em\TeX}}}

\usepackage{graphicx}
\usepackage{subcaption}

\title[Exploring Interaction Patterns for Debugging]{Exploring Interaction Patterns for Debugging: Enhancing Conversational Capabilities of AI-assistants}

\author{Bhavya Chopra}
\authornote{Equal contribution}
\affiliation{
  \institution{Microsoft}
  \country{Bengaluru, India}
  }
\email{t-bhchopra@microsoft.com}

\author{Yasharth Bajpai}
\authornotemark[1]
\affiliation{
  \institution{Microsoft}
  \country{Bengaluru, India}
  }
\email{ybajpai@microsoft.com}

\author{Param Biyani}
\affiliation{
  \institution{Microsoft}
  \country{Bengaluru, India}
  }
\email{t-pbiyani@microsoft.com}

\author{Gustavo Soares}
\affiliation{
  \institution{Microsoft}
  \country{Redmond, Washington, USA}
  }
\email{gustavo.soares@microsoft.com}

\author{Arjun Radhakrishna}
\affiliation{
  \institution{Microsoft}
  \country{Redmond, Washington, USA}
  }
\email{arradha@microsoft.com}

\author{Chris Parnin}
\affiliation{
  \institution{Microsoft}
  \country{Redmond, Washington, USA}
  }
\email{chris.parnin@microsoft.com}

\author{Sumit Gulwani}
\affiliation{
  \institution{Microsoft}
  \country{Redmond, Washington, USA}
  }
\email{sumitg@microsoft.com}

\newcommand{\tool}[0]{\textsc{Robin}}
\newcommand{\baseline}[0]{Baseline}

\newcommand\code[1]{{\texttt{\detokenize{#1}}}}

\begin{abstract}
  The widespread availability of Large Language Models (LLMs) within Integrated Development Environments (IDEs) has led to their speedy adoption. Conversational interactions with LLMs enable programmers to obtain natural language explanations for various software development tasks. However, LLMs often leap to action without sufficient context, giving rise to implicit assumptions and inaccurate responses. Conversations between developers and LLMs are primarily structured as question-answer pairs, where the developer is responsible for asking the the right questions and sustaining conversations across multiple turns. In this paper, we draw inspiration from interaction patterns and conversation analysis---to design \tool{}, an enhanced conversational AI-assistant for debugging. Through a within-subjects user study with 12 industry professionals, we find that equipping the LLM to---(1) leverage the \textit{insert expansion} interaction pattern, (2) facilitate \textit{turn-taking}, and (3) utilize \textit{debugging workflows}---leads to lowered conversation barriers, effective fault localization, and 5x improvement in bug resolution rates.
\end{abstract}

\date{January 2024}



\begin{document}

\maketitle

\section{Introduction}
\label{sec:intro}
The emergence of conversational Large Language Models (LLMs) like ChatGPT~\cite{chatgpt} have commenced an \textit{age of Copilots}~\cite{ageCopilots}, with increasing integration of LLMs into products as chat interfaces. Software developers are increasingly relying on conversational LLMs to express intent, suggest repairs, and receive explanations in natural language \cite{usability_survey_ICSE24}. Interactions between developers and conversational agents like ChatGPT are primarily structured as unilateral \textit{\textbf{question-answer adjacency pairs}}, an \textit{interaction pattern}~\cite{DALE201343} where the developer is responsible for starting the conversation and the AI agent focuses on generating a response to close the conversation. It requires developers to possess the skill to effectively open each turn of the conversation with the right question. Moreover, in a bid to perform the requested task, conversational LLMs may leap to action in the absence of sufficient information, leading to assumptions that could result in inaccurate responses.

Gricean maxims of conversation propose that effective communication is guided by the right amount and orderly presentation of information, truthfulness, and relevance to the context of discussion \cite{gricean_maxims}. In this regard, the tendency of LLMs to follow the question-answer pair interaction pattern and trying to close the conversation in a single-turn irrespective of the task complexity, leads to generic responses with inadequate depth. The lack of structure in generated responses also contributes to conversational challenges, failing to provide a specific solution to the user's query. These limitations highlight the need for further enhancements to the use of conversational patterns by LLMs to take a step towards Gricean maxims.

Beyond question-answer pairs, other useful interaction patterns could significantly enhance the effectiveness and fluidity of conversations in AI systems. One such pattern is \textit{\textbf{insert expansion}}, which allows the speaker to contribute interactionally relevant information to the conversational turn. An example is seen in a customer-server dialogue where clarification is sought \cite{enwiki:1185118449}: \newline
\indent \indent Customer: \textit{``I would like a turkey sandwich, please.''} \newline
\indent \indent Server: \textit{``White or wholegrain?''} \newline
\indent \indent Customer: \textit{``Wholegrain.''} \newline
\indent \indent Server: \textit{ ``Okay.''} \newline
Here, the server uses an insert expansion to accurately fulfill the customer's request. Moreover, studies have identified that for certain tasks, \textit{\textbf{domain-specific task-oriented}} interaction patterns can smooth the flow of the conversation improving the efficiency of the communication~\cite{DALE201343}.

In this paper, we seek to expand the capabilities of LLMs---by engaging in interaction patterns beyond question-answer pairs. To explore the benefits of alternative patterns, we built \tool{}, an enhanced AI-assistant for debugging as an extension within the Visual Studio IDE~\cite{vs}. \tool{} leverages the current state of the conversation, IDE context, and domain knowledge about debugging workflows and strategies to dynamically choose the interaction pattern, alternating between question-answer pairs and insert expansions to create domain-specific interaction patterns for debugging conversations. Additionally, several commercial tools provide follow-ups as an aid to begin or continue conversations \cite{copilotX,tabnine}. We tuned the conversation follow-up suggestions provided by \tool{} to generate questions that are natural and appropriate responses to the agent's message, rather than following the generic question-answer pattern presented in existing Copilots.

Through a within-subjects user study with 12 industry professionals (Section \ref{sec:user_study}), we observe a 5-times higher success rate in task completion with \tool. We find that equipping \tool{} with knowledge of interaction patterns and workflows specific to debugging stimulates a collaborative behaviour, leading to deeper investigation of bugs and correct identification of their root cause. Study participants further reported that \tool{} provides actionable plans to investigate bugs by guiding them to use debugger features within Visual Studio, and lowers conversational barriers with meaningful suggestions for follow-ups. In sections \ref{sec:discussion} and \ref{sec:future-work}, we discuss the potential for personalization and deeper integration of such AI-assistants within IDEs, and the generalizability of this approach to assist developers with other software development tasks.

\section{Background and Related Work}
\label{sec:related_work}

Traditionally, programmers have relied on online tutorials, documentation, and question-answer forums like StackOverflow, and acquired strategies for software development \cite{strategic_programming_knowledge}. Research towards developing conversational agents for software development tasks has been in progress since much longer, with the vision of \textit{Programmer's Apprentice} \cite{programmer_apprentice}, and identification of desirable attributes of pair programming transferable to human-agent conversations \cite{human-agent_pairprogramming_VLHCC}. The emergence of LLMs has made this vision more pragmatic, with their increasing use for code generation (in auto-completion interfaces) and natural language explanations (in chat-based interfaces) \cite{usability_survey_ICSE24}. The evaluation and characterization of developer use of completion models shows developers' positive perceptions of GitHub Copilot, despite minimal differences in task completion time \cite{expectation_vs_experience_CHI22}. Recent work now focuses on facilitating meaningful conversations for software development with LLMs through multi-turn contextual conversations enabled by maintaining transcripts of prior conversations \cite{programmers_assistant_IUI_23}.

Literature in the domain of conversation analysis describes turn-taking and interaction patterns such as \textit{repair}, \textit{adjacency pairs} (such as question-answer, offer-acceptance/refusal, \& compliment-response), and \textit{sequence expansions} (pre-/insert/post-expansion) to be central to human-human interactions~\cite{Conversational_UX_Design_Book, ConversationAnalysis, TurnTaking_SimplestSystematics}. In this realm of conversational LLMs, especially with systems like ChatGPT, GitHub Copilot Chat~\cite{copilotX}, and Tabnine's AI assistant \& Chat~\cite{tabnine}, studying interaction patterns in the context of human-agent interactions can play a crucial role in enabling natural dialog. Recent work by \citet{göldi2023insertexpansions} also uses the insert expansion pattern to assist end-users in receiving preferred responses from the LLM. The work inspired us to investigate the applicability of these interaction patterns in a technical domain such as code debugging. 

Lastly, our work leverages research on debugging practices, strategies, and workflows. Developers tend to debug their code in two primary modes, including (1) identification of the error, and (2) fixing of the error \cite{debugging_techniques_PLATEAU23}.

\section{Enhancing AI-assisted debugging}
\label{sec:debug_assistant}

\begin{figure*}[t!]
    \centering
    \begin{subfigure}[t]{0.48\textwidth}
        \centering
        \includegraphics[width=0.96\textwidth]{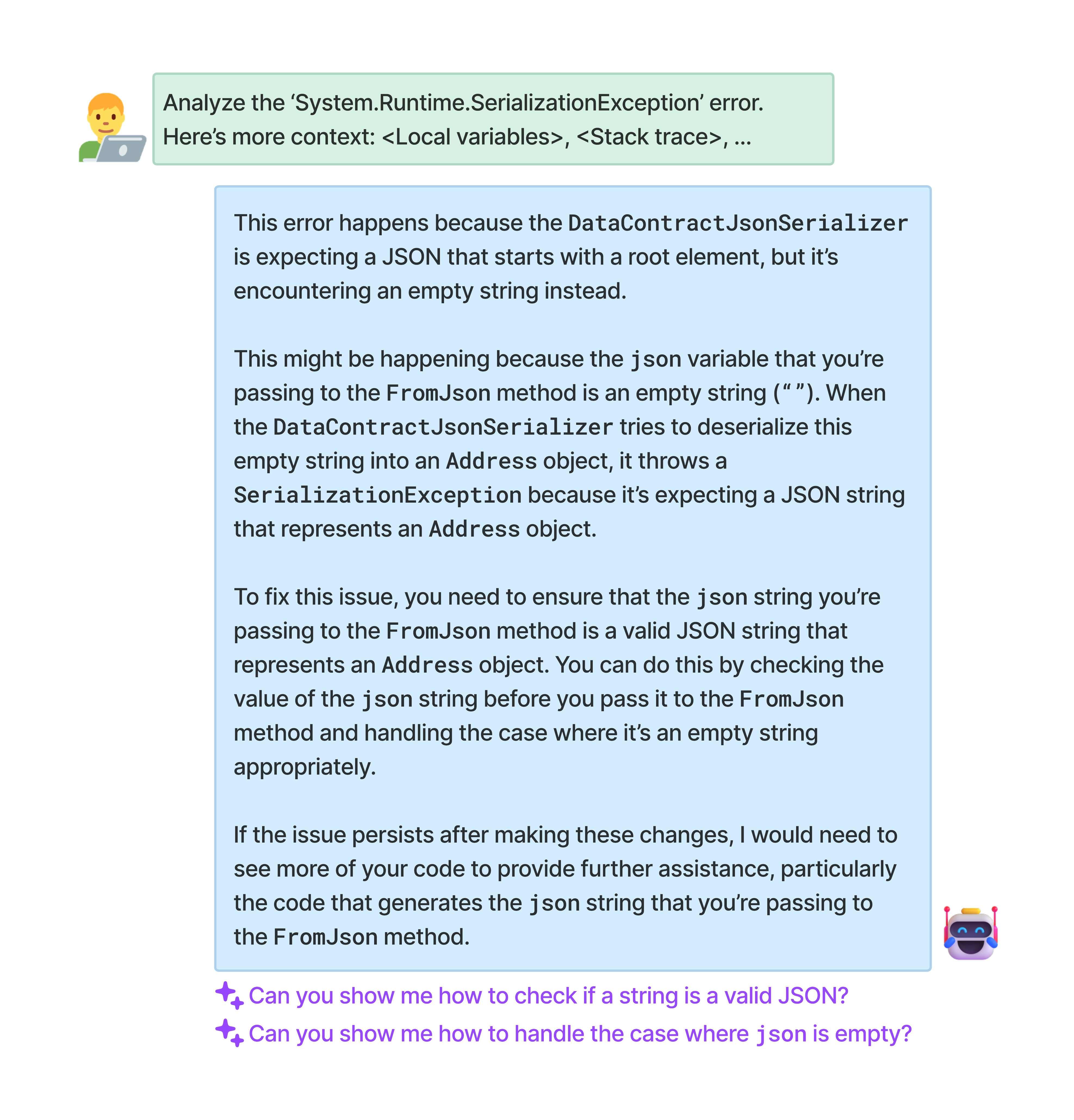}
        \caption{Single-turn response by the baseline assistant}
        \label{fig:bad_response}
    \end{subfigure}%
    ~ 
    \begin{subfigure}[t]{0.48\textwidth}
        \centering
        \includegraphics[width=\textwidth]{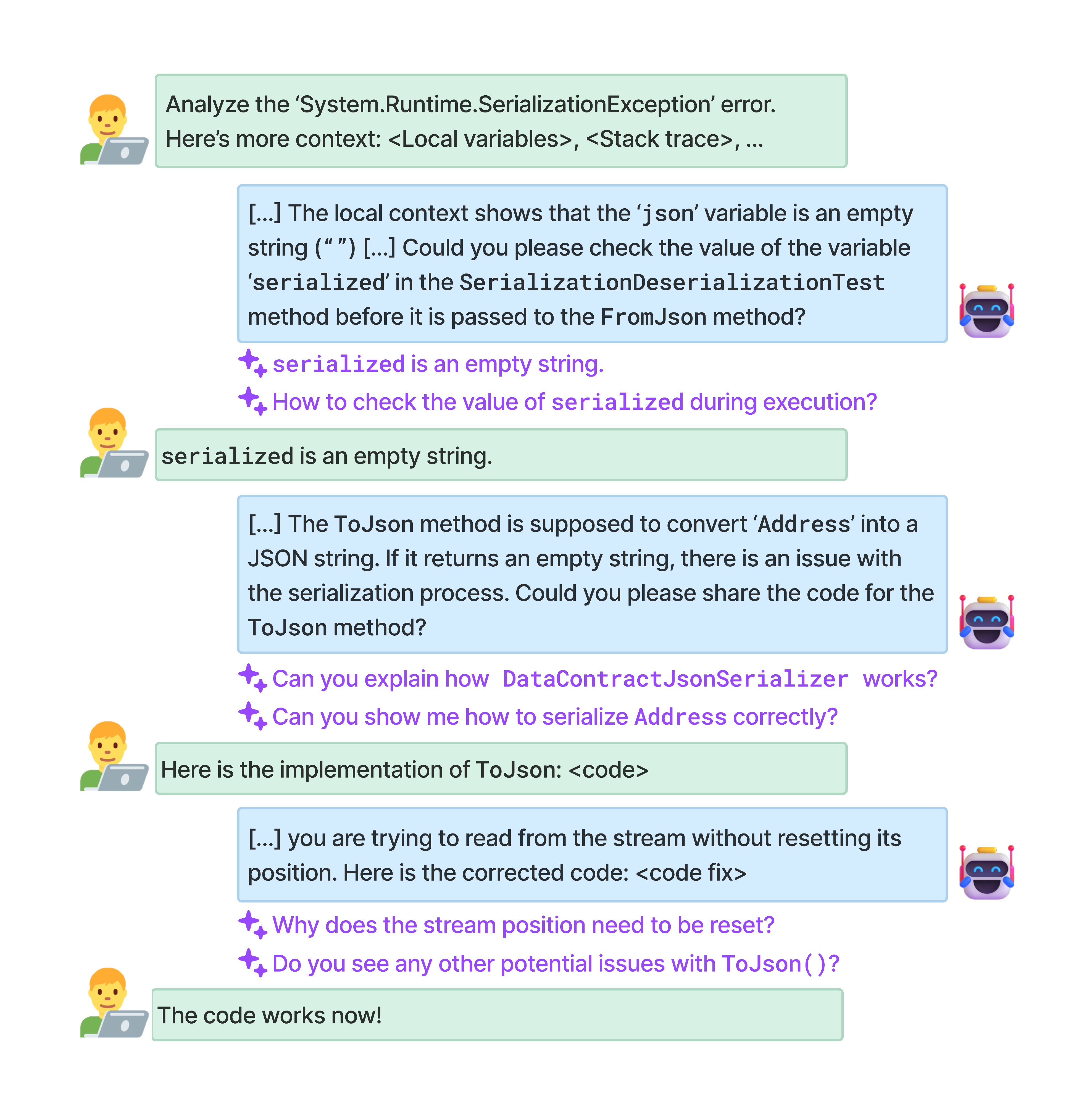}
        \caption{Conversation with \tool}
        \label{fig:good_conversation}
    \end{subfigure}
    \caption{Contrasting conversations for the same task on both the AI-assistants. Text in purple represent the follow-up questions generated by the AI-assistants}
\end{figure*}

Debugging is an essential and time-intensive part of the software development
process. When debugging, developers first identify, localize, understand and then finally fix bugs \cite{debugging_techniques_PLATEAU23}. Unlike other programming tasks, debugging is more free form---debugging assistants must deal with long conversations that involve a wide range of tasks such as setting breakpoints, running or re-running tests, understanding foreign library code, and so on.

To understand the shortcomings of conversational LLMs as debugging assistants, we consider the assistant response in Figure~\ref{fig:bad_response}. This response is generated by a the GitHub Copilot Chat Preview extension for Visual Studio\footnote{\url{https://marketplace.visualstudio.com/items?itemName=VisualStudioExptTeam.VSGitHubCopilot}}, which we refer to as the baseline debugger assistant in this paper. In Figure~\ref{fig:bad_response}, apart from paragraph 1 and 4 that provide general information, the response
consists of $3$ parts that are specific to the code:
(1) a reason for the exception in the current code (paragraph $2$),
(2) a fix (paragraph $3$), and
(3) follow-up utterances (purple links below the paragraph $4$).
Paragraph $2$ starts with the phrase ``This might be happening because\ldots", an indicator that the assistant is not confident in the explanation, and paragraph $3$ proceeds to suggest a fix based on the explanation from paragraph $2$. Here, the explanation is incomplete. While the string being empty is the immediate cause, the root cause is related to why the string was empty at that point in the execution. Due to this superficial diagnosis in paragraph $2$, the suggested fix in paragraph $3$ papers over the symptom and does not address the underlying issue. Ideally, we would like the assistant to first ensure that its diagnosis of the problem is correct before proceeding. However, LLMs are tuned to produce comprehensive single-turn answers, making guesses along the way despite insufficient information. Another problem with the response is the follow-ups generated---these are utterances that the assistant thinks the user might be likely to say in the next turn. Here, the follow-ups generated do not directly continue the conversation, but instead ask new questions. The follow-ups from Figure~\ref{fig:bad_response} prime the user towards
question-answering style conversations. \\

We now describe the key components of \tool{} and how it aims to provide an enhanced debugging experience.

\begin{figure*}[t!]
        \centering
        \includegraphics[width=0.8\linewidth]{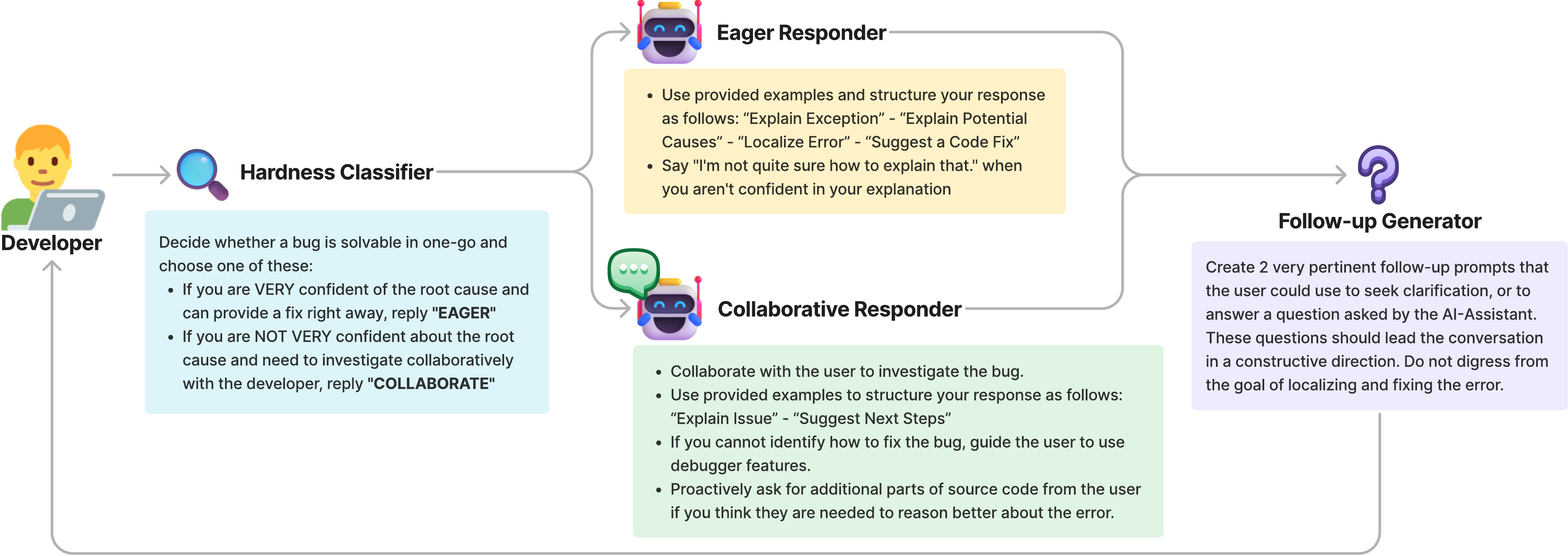}\
        \caption{Agent based workflow for \tool. The \textit{\textbf{Hardness Classifier}} determines the bug severity and decides if it can be resolved in one-shot. The \textit{\textbf{Eager Responder}} suggests a fix in a single-turn response for such errors. The \textit{\textbf{Collaborative Responder}} engages in a multi-turn conversation, providing instructions to use the debugger and seeking additional information from the developer as need arises. The \textit{\textbf{Follow-up Generator}} uses the conversation context to produce prompts that the user will likely say next.}
        \label{fig: arch}
\end{figure*}

\subsubsection*{Insert Expansion for Collaborative Conversations.}
As depicted in Section~\ref{sec:intro}, the \textit{insert expansion} interaction pattern is one where a speaker (in our case, the debugging assistant) does not attempt to answer the other speaker's request immediately, but uses multiple rounds of auxiliary interaction before finally responding to the primary request. In the debugging domain, this conversation pattern can be adapted to aid during the different phases of debugging: fault identification, localization, comprehension, and fixing. While the breadth of possible applications of the insert expansion pattern is large, we go over some common cases.

During the fault identification and localization phases, the assistant can ask for more information from the user. Figure~\ref{fig:good_conversation} shows two examples of insert expansions where the assistant asks the user to provide the value of the variable \code{serialized} and the code for the \code{ToJson} method. Due to  the wealth of debugging-related information available in an IDE, most information cannot be provided to the assistant preemptively due to token limits of language models and other quality concerns. On the other hand, not using the available information leads to vague or incorrect replies. During the fixing phase, the auxiliary interactions aid in collaboratively choosing which of the multiple potential fixes to use. Here, we want the assistant to leverage the user's expertise and familiarity with the code base rather than the assistant making unilateral decisions.
In addition to these, the auxiliary interactions can be used to explain the
steps a user needs to perform during identification and localization phases, to
propose and discuss hypotheses during the fixing phase.

\subsubsection*{Follow-ups for Guided Turn-Taking.}
Follow-ups are aids for easing conversations that are currently being used by several commercial tools~\cite{copilotX, tabnine}. Along with the response, the assistant produces one or more follow-up utterances (purple links below each assistant response in Figure~\ref{fig:bad_response}). These follow-ups serve two purposes. Firstly, in case the utterance exactly matches the user's intent, a user may directly click on it to use it as their next utterance. Second, even in cases where the utterance does not match the user's intent, they serve as a ``style guide'' to direct the user towards the right conversation patterns. Figure~\ref{fig:good_conversation} shows two examples of follow-ups that guide the conversation towards the insert expansion pattern (in the first response by the assistant). The follow-ups continue the conversation and do not just ask related one-off questions like in Figure~\ref{fig:bad_response}. In the first case, the follow-up \emph{``\code{serialized} is an empty string''} is a likely answer to the assistant's question. The second case, \emph{``How to check the value of \code{serialized} during execution?''} is more interesting, showing a possibility of the user opening an additional insert expansion to learn how to find the answer for \tool's question. Further, these follow-ups are both domain- and process-aware. That is, they are apt for the particular circumstances of the current debugging process, specifically related to the investigative nature of the localization phase of debugging.

\subsubsection*{Implementing Conversation Patterns.}
Figure~\ref{fig: arch} shows the high-level overview of the architecture of \tool\, showing the different language-model driven components, along with the salient parts of the prompts used in each. Intuitively, the eager responder and the collaborative responder correspond to the question-answer pair and insert expansion styles of conversations. The hardness classifier decides if a debugging task is straight-forward enough to use the question-answer pair style, or if a collaborative session should be started.

\begin{figure*}[thp]
        \centering
        \includegraphics[width=0.8\linewidth]{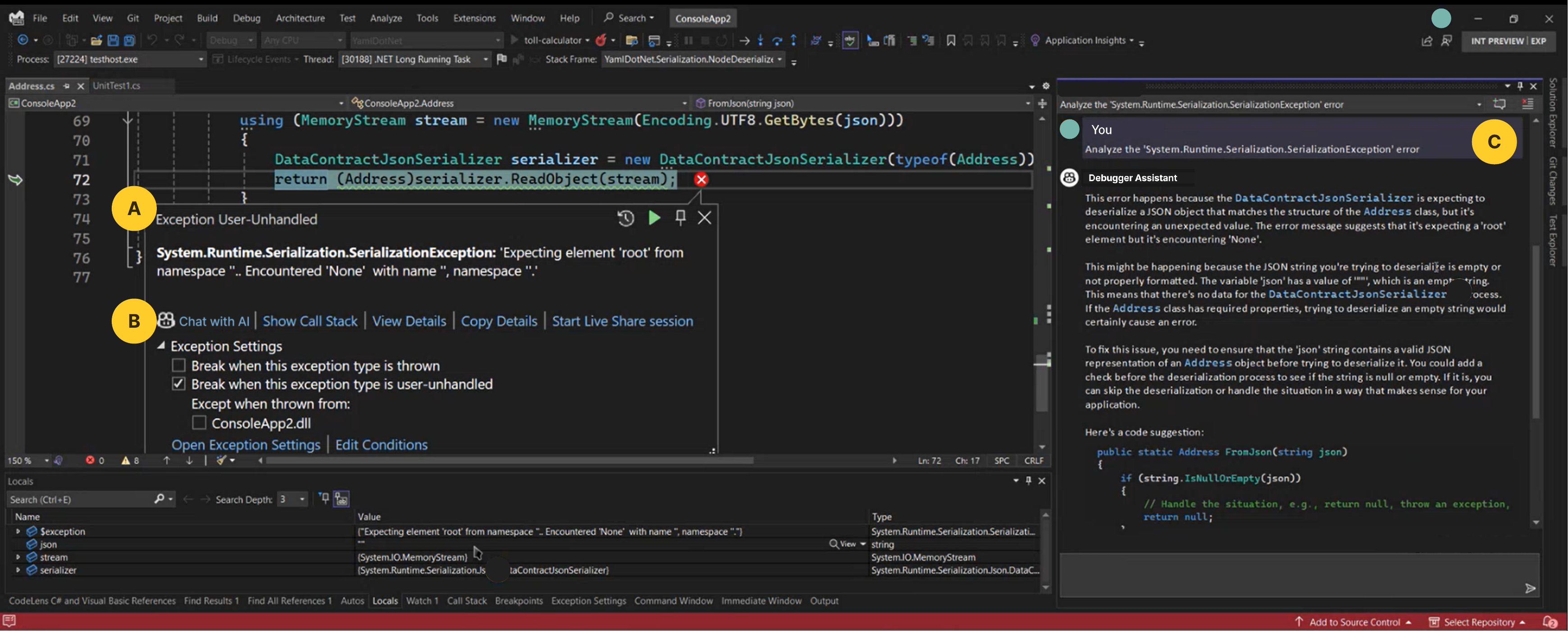}\
        \caption{Study Setup. Participants are exposed to the exception window (A), which has a ``Chat with AI'' button to invoke the AI-assistant (B). The assistant is available as a chat panel on the right side (C).}
        \label{fig:setup}
\end{figure*}

\section{User Study: Evaluating the role of interaction patterns in AI-assisted debugging}
\label{sec:user_study}

We conducted a within-subjects task-based study with 12 software developers to understand if \tool{} can better assist them in successful bug localization and resolution.

\subsection{Methodology}
\subsubsection*{Participants} We recruited participants with at least 1 year of
experience in developing with C\# from our organization through snowball sampling (Table \ref{table:participants}). Prior consent was taken from all participants to audio and screen record our interactions, and use their anonymized quotes for research purposes. 

\begin{table}[]
\caption{Overview of participants, their experience, and self-reported expertise in C\#, alongside the AI-assistants used for each task}
{\small
\scalebox{0.9}{
\begin{tabular}{llllll}
\toprule
\textbf{ID} & \textbf{Exp. (yrs)} & \textbf{C\# expertise} & \textbf{Task 1} & \textbf{Task 2}\\ 
\midrule
P1   & 2.5   & Intermediate  & Baseline & \tool \\
P2   & 1     & Beginner      & Baseline & \tool \\
P3   & 5     & Intermediate  & Baseline & \tool \\
P4   & 6     & Intermediate  & Baseline & \tool \\
P5   & 10    & Expert        & Baseline & \tool \\
P6   & 21    & Expert        & Baseline & \tool \\
P7   & 4     & Intermediate  & \tool  & Baseline \\
P8   & 7     & Expert        & \tool  & Baseline \\
P9   & 2.5   & Intermediate  & \tool  & Baseline \\
P10   & 2.5  & Intermediate  & \tool  & Baseline \\
P11   & 7    & Expert        & \tool  & Baseline \\
P12   & 1    & Beginner      & \tool  & Baseline \\
\bottomrule
\end{tabular}
}}
\label{table:participants}
\end{table}


\subsubsection*{Study protocol} We conducted a within-subjects study using two versions of the AI-assistant integrated within Visual Studio, both of which have access to the exception, call stack and local variables at the point of exception ---
(A) \textit{\textbf{Baseline:}} GitHub Copilot Chat Preview extension, and
(B) \textbf{\tool:} as described in Section \ref{sec:debug_assistant}. 
We conducted one-hour long video conferencing sessions with each of the participants. The study began with demographic questions about their role, programming background in C\#, use of Visual Studio, and prior experience with using LLMs. Next, participants were presented with a warm-up task to familiarize them with the AI-assistant and the development environment. Then, participants were randomly assigned to start with one of the AI-assistants for the first task, and then use the other assistant, for the second task. Based on average task solving times observed in our pilot studies, we added time-bounds of 15 and 25 minutes to tasks 1 and 2, respectively. Finally, after the participants had attempted both tasks, we asked semi-structured questions to seek qualitative feedback.

\subsubsection*{Tasks}
\label{sec:tasks}
We selected one warm-up task and two study tasks that involved comprehending, localizing and fixing run-time exceptions. Figure \ref{fig:setup} shows the task setup for the study. The tasks are comprised of bugs mined from open-source repositories, and adapted for increased tractability. 

\textbf{(Warm-up Task):} An \textit{Index Out Of Range} exception encountered in accessing a list element. This beginner-friendly debugging task helped bring participants up-to-speed with the integrated AI-assistant and available debugging tools in Visual Studio. 

\textbf{(Task 1) Serialization Exception\footnote{\textit{protobuf-net}, issue\# 191: \url{https://github.com/protobuf-net/protobuf-net/issues/191}}:} A moderately difficult task, exposing participants to a \code{Serialization Exception}, arising through the \code{System.Text.JSON} library where it encounters a null character while trying to deserialize a string. The task requires participants to localize the bug to the \code{ToJson} method, and identify a subtle logical error of not resetting the stream position to 0.

\textbf{(Task 2) Arithmetic Overflow Exception\footnote{\textit{YAMLDotNet}, issue\# 673: \url{https://github.com/aaubry/YamlDotNet/issues/673}}:} Participants are exposed to an obscure \code{YAMLException} wrapped around an \code{Arithmetic Overflow Exception}. Bug localization involves finding the correct branch of exploration; and navigating a few call stacks deep to locate the \code{DeserializeIntegerHelper} method. The problem is then reduced to handling a flaw in the logical handling of a lower-edge integer value manipulation.

\subsubsection*{Analysis}
We performed inductive thematic analysis with two annotators to analyze qualitative feedback and on-screen interactions from each session, while noting the timestamps and gathering a log of activities performed by the participants. We also assessed the performance of each AI-assistant based on the time taken by participants to perform the debugging task, and successful identification of the root cause of the bugs.

\begin{figure*}[t]
    \begin{minipage}{0.48\linewidth}
        \includegraphics[width=\linewidth]{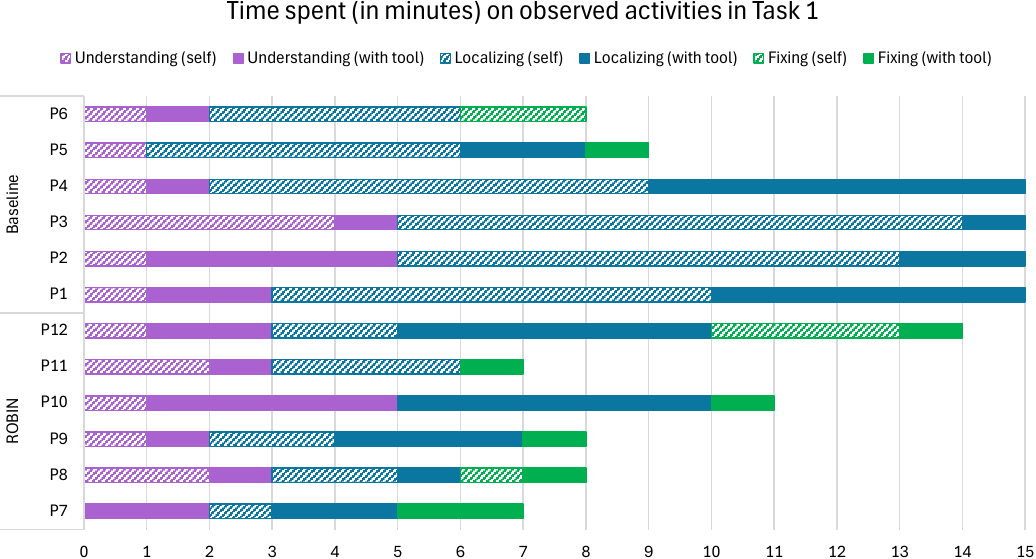}
    \end{minipage}%
    \quad
    \begin{minipage}{0.48\linewidth}
        \includegraphics[width=\linewidth]{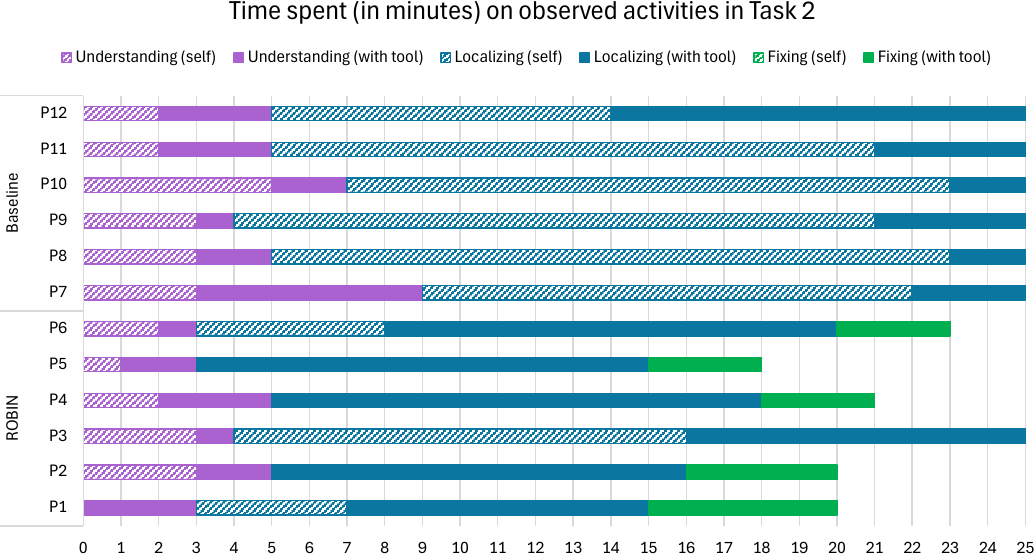}
    \end{minipage}
    \caption{Time spent on each stage of the tasks, with and without AI assistance. Task 1 (Left), Task 2 (Right). Developers spent significantly higher time in localizing errors themselves with the baseline AI-assistant.}
    \label{fig:time-spent-task1}
    \label{fig:time-spent-task2}
\end{figure*}

\begin{table*}[t]
\caption{Overview of task completion rates and engagement with the AI-assistants.}
{\small
\scalebox{0.9}{
\begin{tabular}{llllllll}
\toprule
\textbf{Task}         & {\textbf{\# localizations}} & \textbf{\# fixes}  & {\textbf{\# avg prompts}} & \textbf{\# avg follow-ups used} & \textbf{avg time (with tool)} & \textbf{avg time (self)}\\ 
\midrule
Task 1---\tool     & 6 \textit{of} 6          & 6 \textit{of} 6    & 2 (+53.8\%) & 1.67 (+138.6\%) & 5.67 mins &  3.5 mins\\
Task 1---Baseline & 3 \textit{of} 6         & 2 \textit{of} 6    & 1.3 & 0.7 &  4.3 mins&  8.5 mins\\
\midrule
Task 2---\tool     & 5 \textit{of} 6          & 5 \textit{of} 6    & 3.7 (+85\%) & 3.6 (+990\%) & 15.8 mins & 5.3 mins\\
Task 2---Baseline      & 2 \textit{of} 6          & 0 \textit{of} 6   & 2 & 0.33 & 7.2 mins& 17.8 mins\\
\bottomrule
\end{tabular}}}
\label{table:debugging-behaviours}
\end{table*}

\subsection{Findings}
We provide a summary of participants' interactions with the baseline AI-assistant and \tool{} in section \ref{sec:task_performance}, followed by qualitative insights in sections \ref{sec:premature-localization}--\ref{sec:follow-up-suggestions}. We find that although both assistants are equally helpful in enhancing participant understanding of bugs, \tool~significantly boosts productivity with accurate bug localization and fixes, while enabling learning experiences for developers, owing to its collaborative behaviour.

\subsubsection{Task Performance and Debugging Behaviors}
\label{sec:task_performance}
We logged 104 prompts made by 12 participants in total. Table \ref{table:debugging-behaviours} provides a summary of task completion metrics. Utilizing \tool, we note enhanced task-completion success rates, including a \textbf{2x improvement in bug localization} and a \textbf{5x improvement in bug resolution} ($p<0.05$, Chi-Square test: $13.594$). We also observe that participants' debugging process is accelerated with \tool's assistance. As seen in Figure \ref{fig:time-spent-task1}, there is prolonged self-effort in bug localization with the baseline AI-assistant, while \tool{} hand-holds developers through bug localization and fix generation. Table \ref{table:debugging-behaviours} also indicates an increased engagement with \tool{}, supported by participants spending more time in debugging with \tool{} than on their own, and a significantly greater acceptance rate of the generated follow-up questions.

\subsubsection{Single-turn conversations lead to sub-optimal fixes}
\label{sec:premature-localization}
Most participants did not accept the fixes suggested by the baseline AI-assistant (P2, P5--P12). Across all sessions, participants believed that the baseline assistant only helped them with handling the errors better by throwing custom exceptions or handling case-by-case values using conditional logic. P9 said, \textit{``I would likely not use this fix. I could throw a custom \code{ArgumentException} to handle this better, but I still want to find the deeper issue as to why the string is empty in the first place."}

Few participants further reflected on this characteristic of the baseline debugging assistant, and hypothesized that these sub-optimal fixes are a result of \textit{``pre-mature bug localization''}. P10 said that fixes generated by \baseline\ are trying to \textit{``cure the symptoms instead of the actual disease,''} and this sentiment was shared by P7--P9 as well. P7 said, \textit{``this is trying to localize the bug a bit too soon and just hide it, instead of finding what is giving rise to the issue.''}

On the other hand, participants notice the investigative step-by-step approach followed by \tool{} in locating the root cause of the issue. All participants accepted fixes provided by \tool{} and are satisfied with the accompanying explanation. P5 commented that \textit{``This is providing me thoughtful explanations and steps. The other version gave me code to solve the bug upfront, which confused me even more"}.

\subsubsection{Collaborative debugging equips developers with actionable plans}
\label{sec:collaborative-approach}
Participants appreciated \tool's collaborative approach, where-in it guides on how and where to add breakpoints, and \textit{``telling which variables to keep track of''}:

\begin{quotation}
[...] To confirm this, you can add a breakpoint at this line and check the value of \code{result} when the breakpoint is hit. If \code{result} is \\$9223372036854775808$, then this is where the overflow is happening. Please perform this step and let me know what you find.
\end{quotation}

All participants found value in \tool's responses, and followed steps to add breakpoints and step-through to report variable values. Figure \ref{fig:good_conversation} also shows a snapshot from a conversation where \tool\ requests additional debugging information.

This characteristic often led beginner participants to draw parallels between \tool\ and senior software developers. P2 said, \textit{``I often get stuck with the most pointless bugs while working, and you don't feel like going to ask a senior SDE. This chat assistant would help a lot in these cases.''} Few participants also viewed their interactions with \tool\ as learning experiences. The exploratory nature of responses aided participants' navigation of unfamiliar source code as they solved the study tasks. P8, an expert developer, expressed that such debugging assistance would be \textit{``invaluable to new contributors''}. Further, performing the tasks proposed by \tool{} helped beginner participants in learning features within the IDE, while \textit{``feeling better equipped and knowing what to do the next time I see a similar bug"} (P1).

P11 was not completely convinced to use chat-based assistants for debugging initially, and quoted \textit{``incoherence''} and \textit{``generic responses''} to be significant deterrents. However, their initial experience with using \tool\ for Task 1 changed their perspective to envision how chat-based agents can use \textit{``chain-of-thought''} to appropriately hand-hold developers in debugging tasks. We also note P9's experience, where they expected an actionable plan on which variable values to track from the baseline debugging assistant after having been exposed to \tool{}, and expressed how they felt \textit{``stuck,''} not knowing \textit{``how to get more help''} from the assistant.

\subsubsection{Aligned follow-up suggestions lower conversation barriers}
\label{sec:follow-up-suggestions}
P10 appreciated how \tool's follow-up questions lie in the \textit{``vicinity''} of their query and task at hand. Most participants expressed that \tool{} provided them with the exact question they would want to ask next (P1, P2, P4--P12). These prompts for follow-up with \tool{} facilitated turn-taking with minimal time spent on formulating queries to continue conversing. Figure \ref{fig:good_conversation} shows follow-ups generated by \tool\ for Task 2. Follow-ups generated by the baseline assistant often weren't specific to the participants' local code and conversational context. For instance, P2 received a follow-up suggestion to understand serialization using the \code{JsonSerializer} library, instead of \code{DataContractJsonSerializer} which was used in their code.

\section{Discussion}
\label{sec:discussion}

\subsubsection*{Personalization based on developer experience}
Our study participants belonged to diverse levels of expertise (1--21 years). We learn that developers at different stages in their career have different needs and areas with which the assistant is of more help to them. A \textit{beginner} or \textit{intermediate} user often relied on the understanding of the exception and hypothesis provided by the assistant; however, for an \textit{expert}, both these were often clear from just minor exploration of the exception information. Similarly, the localization assistance needed by a \textit{beginner} was very different in terms of the hand-holding they expected the assistant to provide as compared to an expert who would probably achieve the task without a step-by-step instruction to accomplish routine tasks. Section \ref{sec:collaborative-approach} describes how beginners in the study often perceived \tool\ as an educative tool, comparing the experience to receiving guidance from senior team members. As a next step, we aim to explore ways to align \tool's responses with developer experience and familiarity with code-base.

\subsubsection*{Need for deeper integration with IDE}
\label{sec:deeper-integration}
Participants also expressed a need for both, \tool\ and the baseline assistant, to have greater awareness of the source code to avoid copy-pasting code into the chat. They also desired a deeper integration within the Visual Studio IDE, such that the AI-assistant could automatically perform UI-actions like setting breakpoints and stepping through the code on their behalf. By enabling AI-assistants to perform visible UI-actions, the assistant can also enable learning experiences for beginners like P12, who expressed that \textit{``watching a video tutorial is better than reading chat responses to figure out how to use debugger features in Visual Studio because there are so many menus and elements in this IDE.''} 

\subsubsection*{Towards effective human-agent communication in debugging}
\emph{"All problems lie in absence of a good conversation -- John Niland"} --- This holds true not only for human-human interactions but human-AI interactions as well. Effective communication often guided by Grice's maxims of \textit{\textbf{quantity}}, \textit{\textbf{quantity}}, \textit{\textbf{relevance}} and \textit{\textbf{manner}}~\cite{gricean_maxims}. Here, we reflect on the study insights in the context of Gricean maxims.

\tool's behaviour in coming back to the user for a clarification or for additional information, rather than providing a \textit{premature fix}, neatly embodies the \emph{\textbf{maxim of quality}}. This was universally appreciated by the participants in the user study, who contrasted this with the baseline's tendency to speculate and act on insufficient information~(Section \ref{sec:premature-localization}).

Towards the \emph{\textbf{maxim of quantity}}, \tool{} uses the insert expansion pattern to promote collaborative debugging. Several participants positively commented on how \tool\ proceeds step-by-step, providing only the explanations and actions that are relevant at a given point, as compared to the baseline which tends to provide a single-turn and generic overview and fix upfront. Additionally, \tool\ picks the Eager Responder to address the warm-up task, providing explanations and code fixes upfront---intuitively, the explanation and fix were simple enough for the users to comprehend with a single-turn response. This also demonstrates \tool's adaptability for varying severity of debugging scenarios.

In Section \ref{sec:follow-up-suggestions}, we note a multi-fold increase in the
usage of follow-ups across both tasks in consideration. Most participants could realise a change in the quality of follow-ups as the AI-assistants were swapped across tasks. This, in conjunction with the 5x higher success rate in using \tool\, showcases the integral role played by aligned follow-ups in guiding the conversation. This implicitly guardrails conversations, while also setting the developers' expectations of the agent's capabilities, taking a step towards the
\textit{\textbf{maxim of relevance}} on both ends.

\section{Future Work \& Conclusion}
\label{sec:future-work}

\tool\ is now a part of the GitHub Copilot Chat extension for Visual Studio, and is an ongoing attempt to mitigate some of the immediate conversational issues frequently encountered in the current crop of AI debugging assistants. Currently, we have explored facilitating turn-taking and the insert expansion interaction pattern, which we believe are pivotal in augmenting the user's experience. Inspired by the findings, we plan to explore the applicability of other similar interaction patterns to enhance conversations for debugging. As discussed in Section \ref{sec:discussion}, we aim to experiment with varied needs of expert professionals and support them through \tool. Further, we look forward to enabling deeper integration of the chat assistant with the IDE through automated actions and improved code references. We plan to investigate the impact of domain-specific interaction patterns for project scaffolding, code migration and re-platforming as potential software engineering tasks which are long-running and time intensive, just like debugging.

\bibliographystyle{ACM-Reference-Format}
\bibliography{bibliography}
\end{document}